\documentclass[final,3p,sort&compress,times,twocolumn]{elsarticle}

\usepackage{amssymb,amsmath,amsthm}
\usepackage{mathrsfs,amsmath}	%for curly F symbole
\usepackage{amsfonts}

\usepackage{esvect}		%for vector arrow
\usepackage{mathtools}	%for below packages
\usepackage{diffcoeff}	%for dif eq
\journal{Physics Letters A}

\usepackage{hyperref}  
\usepackage{tabto}  
\usepackage[euler]{textgreek}
 \usepackage{ragged2e}
 \usepackage{graphicx}
\usepackage[table,xcdraw]{xcolor}
\begin{document}

\begin{frontmatter}

\title{Indications of a Fifth Force Coupling to Baryon Number \\ in the Potter Test of the Weak Equivalence Principle}

\author[1]{Megan~H.~McDuffie}
\author[1,2]{Ephraim Fischbach\corref{cor1}}%
\ead{ephraim@purdue.edu}

\cortext[cor1]{Corresponding author}
\author[3,1]{Dennis E. Krause}
\author[2]{John T. Gruenwald}
\author[1]{Michael~J.~Mueterthies}
\author[4,5]{Carol~Y.~Scarlett}
\author[6]{Belvin Freeman}

\address[1]{Department of Physics and Astronomy, Purdue University, West Lafayette, IN 47907 USA}
\address[2]{Snare, Inc., West Lafayette, IN 47906, USA}
\address[3]{Department of Physics, Wabash College, Crawfordsville, IN, 47933 USA}
\address[4]{Department of Physics, Florida A\&M University, Tallahassee, FL 32307 USA}
\address[5]{Orise Fellow, Argonne National Laboratory, Lemont, IL 60439 USA}
\address[6]{Nexus Technologies, Fletcher, NC 28732 USA}

\begin{abstract}

We have reanalyzed data obtained by Potter in a 1923 experiment aimed at testing whether the accelerations of test masses in the Earth's gravitational field are independent of their compositions. Although Potter concludes that the accelerations of his samples compared to a brass standard were individually consistent with a null result, we show that the pattern formed from a combined plot of all of his data suggests the presence of a fifth force coupling to baryon number.
\end{abstract}

\end{frontmatter}

\section{Introduction}

For many years it has been recognized that tests of the weak equivalence principle (WEP) are an important tool in the search for new physics beyond the Standard Model \cite{Fischbach Book,Adelberger,Damour and Donoghue,Jaeckel,Damour,Will,Tino}.  The reasons are twofold:  First, many extensions of the Standard Model introduce new bosons whose virtual exchange leads to new forces.  Second, it is most likely that the couplings of these new bosons to fermions are not all the same \cite{Fayet 1996,Adelberger 2007,Dent,Damour,Kraiselburd}.  This inevitably leads to forces which depend on the compositions of the test bodies \cite{Dobrescu,Fadeev}, in contrast to gravity which is composition-independent.  Hence, finding a violation of the WEP could lead to the discovery of new physics.

Recently it has been argued that in order to test for the presence of a composition-dependent fifth force in a WEP experiment, data are needed from significantly  more than one independent pairs of test samples whose accelerations are being compared \cite{EPF paper}.  Note that early experiments searching for composition-dependent effects in gravity by Newton \cite{Newton}, Bessel \cite{Bessel}, E\"{o}tv\"{o}s \cite{EPF,EPF Book 1,EPF Book 2}, and Potter \cite{Potter,Potter manuscript} used many different samples.  This is in stark contrast to more recent,  high-precision  WEP experiments \cite{Dicke,Braginskii,Wagner,Asenbaum,Touboul PRL,Touboul CQG}  which typically  compare accelerations involving only two samples.  The reason that early researchers used many different samples is just as relevant today as it was for Newton: One must use a sufficient number of different materials to discern a {\em pattern} if one exists.  Since many different theoretical models exist for new forces, with different couplings and spatial dependences, it may be difficult to predict {\em a priori} what may actually be observed in a given experimental setup.  However, the clearest signal would be the observation of a {\em pattern} in the data, where none should exist.  It is then remarkable that the most precise WEP test utilizing a significant number of samples, the torsion balance experiments by E\"{o}tv\"{o}s, Pek\'{a}r, and Fekete (EPF) \cite{EPF,EPF Book 1,EPF Book 2}, in fact reveal such a pattern.

Motivated by anomalous experimental results and theoretical considerations, Fischbach and colleagues examined the EPF data for an acceleration dependence on baryon number \cite{Fischbach PRL,AoP,Fischbach memoir,Franklin}.  They found that the difference in accelerations between samples was directly proportional to the difference in baryon-to-mass ratios of the samples.  While their proposal to explain these results (a new ``fifth force'' arising from a Yukawa potential proportional to baryon number with a range on the order of several hundred meters) was subsequently excluded by many experiments \cite{Wagner,Tino, Franklin, Fischbach Book, Adelberger, Will}, the {\em pattern} in the EPF data remains unexplained.  In order to resolve this  {\em E\"{o}tv\"{o}s paradox} \cite{Fischbach PoS}, reconciling the EPF data with the lack of confirmation from precise experiments, it has been proposed recently that a new set of WEP experiments should be carried out using a sufficient number of samples, with widely varying baryon-to-mass ratios, to confirm or exclude the pattern observed in the EPF data~\cite{EPF paper}.

It is against this backdrop that we decided to examine the data from the WEP experiment conducted by Potter published in 1923 \cite{Potter}. It is interesting to note that Potter's motivation was not dissimilar to what is motivating modern researchers.  At the time, it was thought that the constituents of nuclei were hydrogen nuclei (protons) or helium nuclei. (The neutron had not yet been discovered.)  Potter wished to see if gravity affected the hydrogen and helium components differently, and so chose a number of samples with different numbers of these more fundamental nuclei.  His experiment, which will be described in more detail below, utilized a pendulum and was less precise than the earlier EPF torsional balance experiment.  From the measured periods of each sample pendulum, Potter calculated and compared the respective accelerations. In his published paper \cite{Potter}, specifically, he reported a null result (each acceleration is within the limit of uncertainty to one another) which suggested the data were consistent with the WEP within experimental uncertainties.

In our reanalysis of the Potter data described below, we observe an unexpected result.  Specifically, we find that Potter's period data, in fact, exhibit a clear dependence on the samples' baryon-to-mass ratios in a manner not unlike that found in the EPF data.  Our discussion of the Potter experiment describes his procedure and presents his results.  We then carry out our reanalysis of the data and detail the evidence for a period dependence on baryon number.  A phenomenological force which could explain these data, and the similar EPF results, is  suggested.  We also consider alternative couplings to atomic number $Z$ and neutron number $N$.   We conclude by discussing the implications of these results and how they might provide an important clue in resolving  the E\"{o}tv\"{o}s paradox.

\section{Potter Experiment Description}
\subsection{Experimental Design and Procedure}

In contrast to the EPF \cite{EPF paper} and Eöt-Wash \cite{Adelberger} torsion balance experiments, the Potter experiment  \cite{Potter} was performed  with a pendulum, improving upon the method used by Bessel \cite{Bessel}.  Since he was not interested in an absolute measurement of gravity, Potter compared the sample pendulum periods with a standard made of brass. Five groups of measurements were performed and the results are presented in Table \ref{Potter_table}.   (We note  that many details describing the experiment were omitted in the published version of Potter's paper, but are included in the original submitted manuscript, which exists in the archives of the Royal Society \cite{Potter manuscript}.)

In the beginning of each group of experiments, the apparatus was calibrated by comparing a brass sample to the brass standard. Both the standard pendulum and test pendulum were of the same composition, and this allowed a calibrated acceleration to be recorded for the remainder of the samples in that group. The ``calibrated'' acceleration was defined as $g$. This calibration technique also provided the period of a brass sample within each group to which the other materials could be compared.

The period, $T_S$, of the standard was measured using a standard clock and found to be 1.7703 seconds. Each of the samples was compared to the period of the standard in the following manner: The sample and the brass standard were released together, and hence were in phase initially. After drifting out of phase, they returned to being in phase, and the total time of this cycle was recorded.  More precision was obtained by timing 100 of these ``beat'' periods.

\subsection{Experiment Groups}

Each group had a distinct experimental setup for the sample pendulum; however, each group was always compared to the standard pendulum by using the period of coincidence in the technique described previously.

Group 1: The first group of measurements determined the periods of brass, followed by lead and steel. This group of samples was measured using a Cambridge and Paul Brass Cylinder. A brass filled bob was measured three times and averaged to calculate and calibrate the period of the brass sample for Group 1. 

Group 2: The second group of samples were studied using a Hilger Brass Cylinder. The calibration was defined by measurements of a brass disc in a Hilger brass cylinder. After the first sample, ammonium fluoride, the pendulum needed to be cleaned and taken apart. When the pendulum was reassembled, a new brass measurement was needed, and the remaining samples were defined as a separate group (Group 3). 

Group 3: Two new brass calibration measurements were made. The calibration was defined by measurements of a brass disc in a Hilger brass cylinder, as well as a brass ring. The period of the brass disc and ring were averaged and recorded. The periods of bismuth, paraffin wax, and duralumin were then independently measured in the same Hilger brass cylinder. 

Group 4: The fourth group of samples was measured in a Hilger duralumin cylinder. A brass disc was used to calibrate the group. After this calibration was carried out, the periods of paraffin wax and mahogany were measured. 

Group 5: The final group of samples was again measured in a Hilger duralumin cylinder, and the calibration was done using a brass disc (similar to the previous group of samples). The only period to be measured in this group was that of paraffin wax. 

\begin{table}[t]
\caption{Results from the Potter experiments \cite{Potter}. $T_S$ is defined as the period of the brass standard.}
\begin{tabular}{lrr}\hline
Material & \multicolumn{1}{c}{Period} & Acceleration \\ \hline\hline
\textit{Group 1--} &                 &                       \\
Brass             & 1.0004172 $T_S$ & $g$                   \\
Lead              & 1.0004193 $T_S$ & 0.999992$g$           \\
Steel             & 1.0004220 $T_S$ & 0.999980$g$           \\ \hline
\textit{Group 2--} &                 &                       \\
Brass             & .9999185 $T_S$  & $g$                   \\
Ammonium fluoride  & .9999175 $T_S$  & 1.000005$g$           \\ \hline
\textit{Group 3--} &                 &                       \\
Brass             & .9999885 $T_S$  & $g$                   \\
Bismuth           & .9999895 $T_S$  & 0.999994$g$           \\
Paraffin wax      & .9999865 $T_S$  & 1.000012$g$           \\
Duralumin         & .9999905 $T_S$  & 0.999992$g$           \\ \hline
\textit{Group 4--} &                 &                       \\
Brass             & 1.0000520 $T_S$ & $g$                   \\
Paraffin wax      & 1.0000490 $T_S$ & 1.000014$g$           \\
Mahogany          & 1.0000485 $T_S$ & 1.000015$g$           \\ \hline
\textit{Group 5--} &                 &                       \\
Brass             & .9999365 $T_S$  & $g$                   \\
Paraffin wax      & .9999335 $T_S$  & 1.000014$g$           \\ \hline
\end{tabular}
\label{Potter_table}
\end{table}

\subsection{Corrections for Systematic Effects}

The values of the periods recorded in Table \ref{Potter_table} are not the raw measured values, but include multiplicative correction factors to account for various systematic effects, which Potter took into consideration. These systematic effects include:

\begin{itemize}
\item A reduction factor was incorporated into the period of the pendulum to correct for a center of mass which was not in the center of the cylinder.

\item Potter also included a correction factor for finite amplitudes estimated to be extremely small. 

\item A buoyancy correction was included for each period, which also included any corrections from temperature and pressure changes.

\item Potter concluded that no correction due to damping was necessary.  

\item Potter used a micrometer to measure the total length of the pendulum arm, since the period depended on the length of the pendulum. He plotted the micrometer reading versus the mass of the loaded substance. Potter then made a correction which is dependent on the amount of deviation from the expected straight line. 

\item Lastly, Potter discussed the correction made for non-rigid motion between the pendulum bob and the wire, as well as between the wire and the knife-edge mounting. Potter calculated a multiplicative correction for the non-rigidity of the connection of the bob and wire, but stated that the correction between the wire and knife mounting was negligible. 
\end{itemize}

\subsection{Potter's Conclusions}

After obtaining the corrected periods for each experiment group, Potter then calculated a value of the gravitational acceleration for each of the samples relative to the brass sample of the experimental group, as shown in Table~\ref{Potter_table}.  Potter observed that, except for the steel sample, none of the periods deviated from the corresponding brass samples by larger than one part in $3\times 10^{5}$.  He concluded that no  differences  in the accelerations  by his samples  beyond the experimental errors were observed.  Unfortunately, he did not include a detailed analysis of his statistical or systematic uncertainties.
 
 Potter was well aware that the sensitivity of his apparatus was much less than the torsion balance used by  EPF, whose work was published shortly before his paper \cite{EPF}.   He mentioned  that he attempted to obtain funding for an E\"otv\"os balance, but failed.  Interestingly, a few years later, he managed to obtain a grant to fund a new experiment using an E\"otv\"os balance.  He published a brief note stating that he again observed no differences in the mass to weight ratios of his samples, this time to one part in $1.5 \times 10^{7}$, but gave no specific results or details \cite{Potter2}.

 \section{Reanalysis of the Potter Data}
 
 \subsection{Motivation}
 
 As in the case of the Potter experiment, it was conventionally accepted that the EPF experiment gave no evidence of a violation of the WEP.  However, as discussed in the Introduction, a reanalysis of the EPF experimental data by Fischbach et al., revealed a pattern in the data which could be explained by a new force coupling to baryon number.  As emphasized in Ref.~\cite{EPF paper}, it is the {\em pattern} in the data which is important, a pattern which should not exist if the WEP is valid.
 
With this in mind, we decided to undertake a reanalysis of the Potter experiment since it used a significant number of samples, and the published paper \cite{Potter} (and unpublished manuscript \cite{Potter manuscript}) contained sufficient details of the experimental procedure and analysis to undertake a reanalysis.  Because the Potter experiment was significantly less sensitive than the EPF experiment, one does not expect to find a pattern in the Potter data, even if a fifth force coupling to baryon number existed, and had been detected by the EPF experiment.

\subsection{Period Values for the Potter Samples}

In our reanalysis, we decided to focus on the periods obtained by Potter for each sample (Table \ref{Potter_table}) rather than the accelerations.  Surprisingly,  Potter did not provide any details how that latter were calculated, and it was not obvious how these results were obtained.

Because of the different setups for each of the experiment groups, the  periods for each sample ($T_i$) must be compared to the  period of brass in each respective group ($T_b$), 
\begin{equation}
\Delta T_{ib} \equiv T_i - T_{b}
\label{Tib}
\end{equation}
where $T_i$ and $T_b$ are both in units of $T_S$, the period of the standard.  Table \ref{Potter_results} presents a summary of the calculated difference in period for each substance compared to brass ($\Delta T_{ib}$).

\begin{table}[]
\caption{Summary of Data for Potter Samples.  Here $\Delta T_{ib}$ is measured in units of $T_{S}$.  Because of the uncertainty in the composition-dependence of mahogany and the difficulty in estimating its value of $B/\mu$, we have not included its value of $\Delta(B/\mu)_{ib}$.   }
\small
\begin{tabular}{lcrr}\hline
Material    & Index ($i$) & $ 10^6 \Delta T_{ib}$ & $10^3 \Delta(B/\mu)_{ib}$ \\ \hline
Lead $-$ Brass         & 1        & 2.100            & $-$1.00121             \\ 
Steel $-$ Brass       & 2        & 4.800            & 0.04672              \\ 
NH$_4$F $-$ Brass      & 3        & $-$1.000           & $-$2.00941             \\ 
Bismuth $-$ Brass     & 4        & 1.000            & $-$1.02305             \\ 
Paraffin wax $-$ Brass & 5        & $-$2.667           & $-$2.24491             \\ 
Duralumin $-$ Brass   & 6        & 2.000            & $-$0.38021             \\ 
Mahogany $-$ Brass    & 7        & $-$3.500           &        \multicolumn{1}{c}{\,\,\,\,\,\,\,\,\,---}        \\ \hline
\end{tabular}
\label{Potter_results}
\end{table}

\subsection{$B/\mu$ Values for the Potter Samples}

Since the EPF experimental data exhibit an acceleration dependence on the baryon-to-mass ratio $B/\mu$, where $\mu = m/m_{\rm H}$ is the mass of the sample in units of the mass of hydrogen $m_{\rm H}$, we calculated  $B/\mu$ for each of the samples used by Potter.  Several of the samples were chosen  for their relatively large hydrogen contribution, as discussed by Potter \cite{Potter}.

Here are some comments on these samples:

\begin{itemize}
\item Paraffin wax (CH$_2$)$_n$ is an obvious choice, given its relatively large hydrogen concentration, and general availability ($B/\mu = 1.006698$).

\item Similar remarks apply to ammonium fluoride, NH$_4$F ($B/\mu = 1.006933$). 

\item The chemical composition of mahogany, unlike those of the other Potter samples whose compositions are well-determined, is uncertain and dependent on its source \cite{Mahogany}.  Therefore, we have not included the mahogany sample in our analysis.

\item Bi and Pb are elements whose $B/\mu$ values are given in Table 2.1 of Ref. \cite{Fischbach Book} (Bi: $B/\mu = 1.007920$ and Pb: $B/\mu = 1.007942$). Although the concept of baryon number ($B$), and hence of $B/\mu$, did not exist at the time of the experiment, Pb($Z=82$) and Bi($Z=83$) are neighbors in the Periodic Table of Elements, and hence have very similar values of $B/\mu$. 

\item From Ref. \cite{Duralumin} the alloy duralumin is composed of Al(94.5\%) + Cu(5\%) + Mg(0.5\%),  by weight. Since Al and Mg are also neighbors in the Periodic Table, any variations in the $B/\mu$ values for duralumin are likely to be small and will be solely due to the Cu content, which is any case is also relatively small. The composition of this alloy of Duralumin yields $B/\mu = 1.008563$. 

\item With respect to steel, there are many alloys of Fe which could be used as ``steel.'' As a representative ``steel'' we use the value quoted in Ref. \cite{Duralumin} for stainless steel whose composition by weight is: Fe (80.6\%), C (0.4\%), Cr (18.0\%), Ni (1.0\%). The composition of this alloy of steel gives $B/\mu~=~1.008990$.

\item Brass is an alloy of copper and zinc, and can have multiple compositions. The composition of brass used in this analysis is quoted from Ref. \cite{Duralumin} to be Cu (60.0\%), Zn (39.0\%), Sn (1.0\%) by weight. The resulting baryon-to-mass ratio for this composition is 1.008943.

\end{itemize}

Table \ref{Potter_results} shows the composition difference of each sample, $i$, compared to brass, where
\begin{equation}
\Delta \left(\frac{B}{\mu}\right)_{ib}  \equiv \left(\frac{B}{\mu}\right)_i - \left(\frac{B}{\mu}\right)_{b} .
\label{Bib}
\end{equation}

\section{Fifth Force Model}

Before continuing, and to motivate our subsequent analysis, let us  examine the effect of a new fifth force on the Potter experiment.  As discussed in Ref.~\cite{EPF paper}, the pattern observed in the EPF experiment can be explained by a force which couples linearly to baryon number,
\begin{equation}
\vec{F}_5 = B\vec{\mathbb{F}}_5
\label{F5}
\end{equation}
where $\vec{\mathbb{F}}_5$ is a constant  force field.  Since the magnitude of this force is much less than the gravitational force, its effect on the Potter experiment can be investigated using a simple pendulum model.

\begin{figure} [h!]
\centering\includegraphics[width=0.3\textwidth]{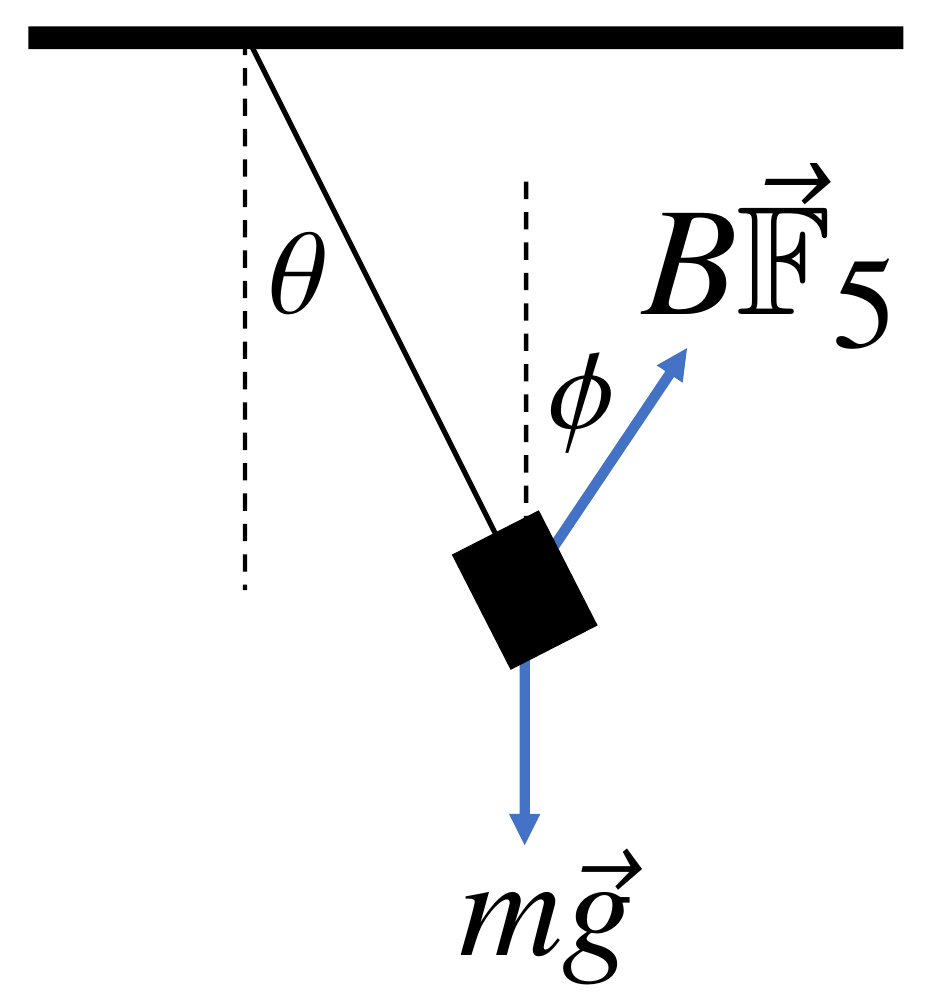}
\caption{Cartoon of a simple pendulum acted upon by the gravitational force ($m\vec g$) and a  5th-force proportional to baryon number    $(B\vec{\mathbb{F}}_5)$.}
\label{pendulum}
\end{figure}

Consider the setup shown in Fig.~\ref{pendulum}.  A small test mass $m$ is suspended from massless string of length $L$, and acted upon by the gravitational force $m\vec{g}$ and a fifth force given by Eq.~(\ref{F5}) which, for simplicity, acts in the same plane as the string tension and the gravitational force.  The tension and fifth force act at angles $\theta$ and $\phi$, respectively, relative to the vertical defined by the local value of $\vec{g}$ as shown.

The torque $\tau$ acting on the pendulum mass is given by (for $\theta \ll 1$), 
\begin{align}
\tau& = -mgL\sin\theta + B{\mathbb{F}}_5L\sin(\theta + \phi) \nonumber \\
	&=-mgL\sin\theta + B{\mathbb{F}}_5L(\sin\theta \cos\phi + \cos\theta \sin\phi)  \nonumber \\
	&\approx (-mg + B{\mathbb{F}}_5 \cos\phi)L\theta + B{\mathbb{F}}_5L\sin\phi.
\label{torqueArb}
\end{align}
Solving for the equilibrium angle ($\tau = 0$), and assuming that the gravitational force is much larger than the fifth force contribution ($mg \gg B{\mathbb{F}_5} \cos\phi$), the equilibrium angle simplifies to,
\begin{equation}
\theta_{\rm eq} =  \frac{B{\mathbb{F}_5}\sin\phi}{(mg - B{\mathbb{F}_5} \cos\phi)} \approx \frac{B{\mathbb{F}_5}\sin\phi}{mg} .
\end{equation}
The equation of motion of a pendulum with the fifth force is then, 
\begin{equation}
\diff[2]{\theta'}{t} = \left(-\frac{g}{L} + \frac{B}{m}\frac{{\mathbb{F}_5 \cos\phi}}{L}\right)\theta' ,
\label{difeq}
\end{equation}
where $\theta' = \theta - \theta_{\rm eq}$ is the angle of the pendulum measured relative to the shifted equilibrium position, rather than from the vertical. Eq.~(\ref{difeq}) is  the  differential equation of a simple pendulum 
\begin{equation}
\diff[2]{\theta'}{t} = -\left(\frac{g_{\rm eff}}{L}\right)\theta,
\label{simpledifeq}
\end{equation}
where the usual value of $g$ is replaced by $g_{\rm eff}$ given by
\begin{equation}
g_{\rm eff} \equiv g - \frac{B}{m} {\mathbb{F}_5 \cos\phi} .
\end{equation}
The period $T$ of the pendulum is then given by
\begin{align}
T &= 2\pi \sqrt{\frac{L}{g_{\rm eff}}} = 
2\pi \sqrt{\frac{L}{g}} \left( 1 - \frac{B}{m} \frac{{\mathbb{F}_5 \cos\phi}}{g}\right)^{-1/2}.
\label{periodEq.}
\end{align}
If we rewrite  the mass in terms of the mass of hydrogen,  $m = \mu m_{\rm H}$, where $m_{\rm H} = m(\textsubscript{1}H^1) = 1.0078251$u, and use
\begin{equation}
\frac{B}{m} \left(\frac{\mathbb{F}_5 \cos\phi}{g}\right) = \frac{B}{\mu}\left(\frac{\mathbb{F}_5 \cos\phi}{m_{\rm H}g} \right) \ll 1, 
\end{equation}
 it follows that 
\begin{equation}
T \approx 2\pi \sqrt{\frac{L}{g}} \left( 1 + \frac{B}{\mu} \frac{{\mathbb{F}_5 \cos\phi}}{2m_{\rm H}g}\right) = T_0 + T_5 .
\label{periodfinal}
\end{equation}
Here the expression for the period has been explicitly separated into the gravitational and fifth-force contributions, $T_0$ and $T_5$, respectively, where, 
\begin{align}
T_0  &\equiv 2\pi \sqrt{\frac{L}{g}}, \\
T_5 & \equiv T_0\left(\frac{{\mathbb{F}_5 \cos\phi}}{2m_{\rm H}g}\right) \left(\frac{B}{\mu}\right).
\end{align}

When computing the difference in the periods of each material ($T_i$) and brass ($T_b$) in the Potter experiment, the difference ($T_i - T_b$) will subtract out the constant gravitational contribution ($T_0$) common to both. This leaves the fifth-force contribution which depends on the baryon-to-mass difference of the two samples, $i$ and $b$, 
\begin{align}
\Delta T_{ib} & = T_i - T_b = T_{5,i} - T_{5,b} \nonumber \\
	&= T_0\left(\frac{{\mathbb{F}_5 \cos\phi}}{2m_{\rm H}g}\right) \left(\frac{B_i}{\mu_i} - \frac{B_b}{\mu_b}\right)\nonumber \\
	&= T_0\left(\frac{{\mathbb{F}_5 \cos\phi}}{2m_{\rm H}g}\right) \Delta \left(\frac{B}{\mu}\right)_{ib}.
\end{align}
Dividing by $T_0$ we find, 
\begin{equation}
\frac{\Delta T_{ib}}{T_0} = \left(\frac{{\mathbb{F}_5 \cos\phi}}{2m_{\rm H}g}\right) \Delta \left(\frac{B}{\mu}\right)_{ib} .
\label{GeneralPeriod}
\end{equation}

As can be seen by this analysis, the intrinsic simplicity of the Potter experimental setup makes it easier to extract directional information of $\vec{F}_{5}$.  While a torsion balance is experiment is most sensitive to horizontal forces, a pendulum is most sensitive to forces acting vertically.  For example, if the vertical component of the fifth force is upward ($|\phi| < \pi/2$),  $T_{5} > 0$, while a downward component  ($|\phi| > \pi/2$) gives $T_{5} < 0$.
For example, if we assume $\vec{\mathbb{F}}_5$ is antiparallel to $\vec{g}$ ($\phi = 0$), it then follows from Eq. (\ref{GeneralPeriod}) that the period of the Potter data is, 
\begin{equation}
\frac{\Delta T_{ib}}{T_0} = \left(\frac{{\mathbb{F}_5}}{2m_{\rm H}g}\right)_{\phi = 0} \Delta \left(\frac{B}{\mu}\right)_{ib} .
\label{ExamplePeriod}
\end{equation}
 If $B_i/\mu_i > B_b/\mu_b$, then  Eq. (\ref{ExamplePeriod}) gives $T_i > T_b$. This is intuitively what we expect from a  force  acting against gravity: The larger the value of $B/\mu$, the greater the  force \emph{opposing} gravity, and hence the lower the net acceleration driving the pendulum.  This then results in a longer period,  $T_i > T_b$.   
 
 These results tell us that if $\Delta T_{ib}$ is plotted versus $\Delta(B/\mu)_{ib}$, and this simple model is correct, we should expect the values to form a line with slope given by $\mathbb{F}_5\cos\phi/2m_{\rm H}g$, and the intercept should vanish.

\section{Plotting the Potter Data}

\subsection{Coupling to Baryon Number}
\begin{figure} [t]
\includegraphics[width=\columnwidth]{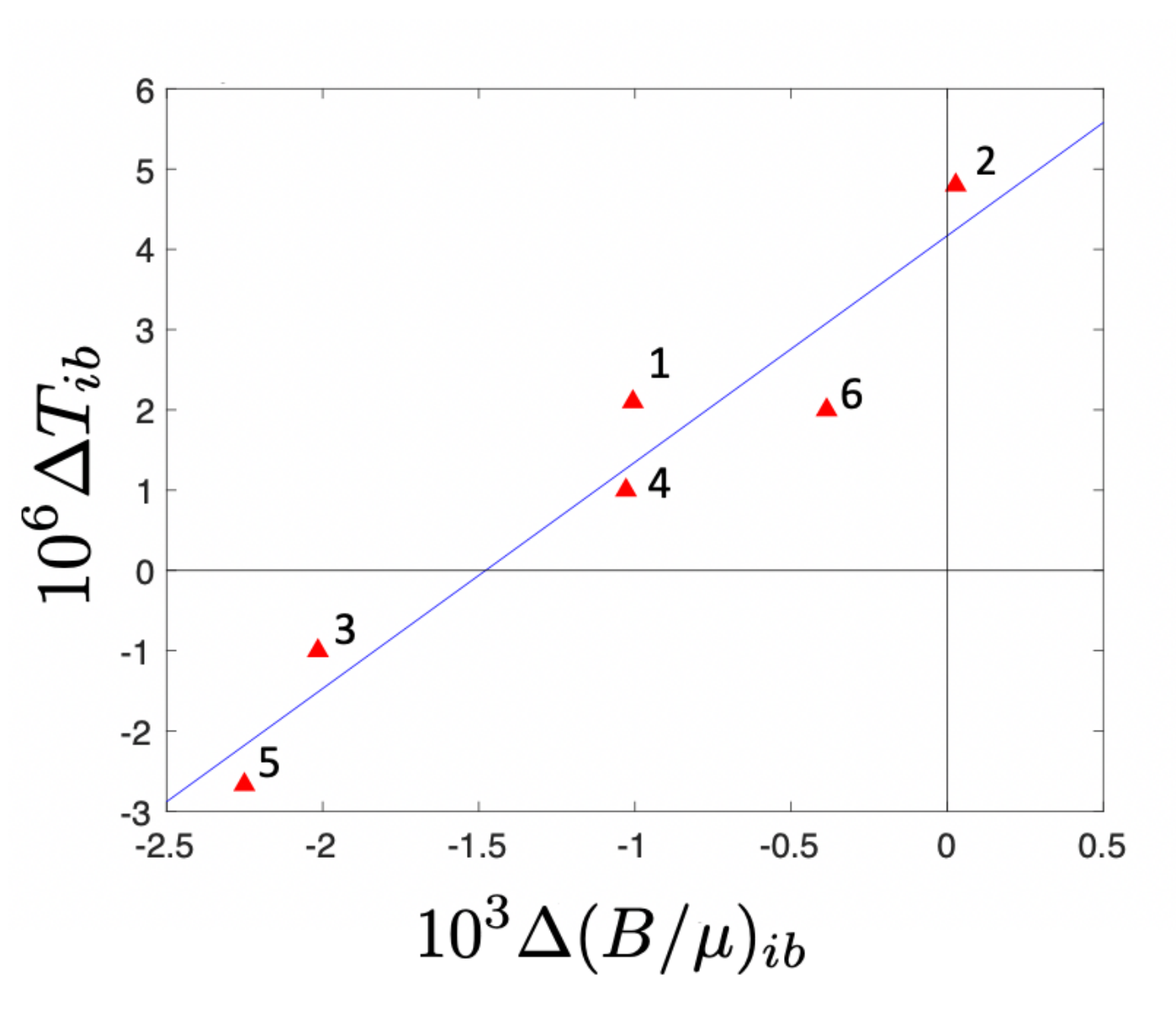}
\caption{Plot of the 6 samples measured by Potter compared to brass. $\Delta T_{ib}$ is defined in Eq. (\ref{Tib}) and $\Delta(B/\mu)_{ib}$ is defined in Eq. (\ref{Bib}), respectively. The 6 samples are tabulated in Table \ref{Potter_results}.}
\label{6points}
\end{figure}

In what follows, we present two figures exhibiting Potter's data. Figure \ref{6points} plots the periods $\Delta T_{ib}$ vs. $\Delta (B/\mu)_{ib}$, the corresponding difference in $B/\mu$ values between sample $i$ and the brass sample for each group. Following the model given by Eq. (\ref{GeneralPeriod}), a linear fit to the data provides the slope and intercept for Fig. \ref{6points}. Given the uncertainty in the composition of mahogany \cite{Mahogany}, that datum has not been included.

Of central importance in the ensuing discussion is that no experimental uncertainties are shown for the data points, since no quantitative uncertainties are actually quoted by Potter for the individual data points. What we have from Potter is the following statement in his summary, dealing with the accelerations of his various samples compared to brass: ``...no difference greater than that attributable to experimental error has been found.'' Stated another way, the acceleration differences, $\Delta (a_{ib}/g)$, were consistent with zero for all of his samples. It follows that the uncertainties of the period differences would also be consistent with zero. Nonetheless, the data evidently suggest a pattern which depends on the composition-dependent quantities $\Delta(B/\mu)_{ib}$. The nonzero slope of the resulting line in Fig. \ref{6points}, is consistent with what would be expected from a composition-dependent fifth force proportional to baryon number. Although the statistical significance of the Potter results cannot be assessed, the ``Potter effect" illustrates that a series of measurements, each consistent with a null result, can nonetheless support an interesting nonzero effect. 

Since there are no uncertainties for $\Delta(B/\mu)_{ib}$, and none quoted for $\Delta T_{ib}$, a simple linear regression is sufficient to find the slope $a_{B/\mu}$, intercept $b_{B/\mu}$, and the coefficient of determination $(R^2)$ of the line in Fig. \ref{6points}:
\begin{subequations}
\begin{align}
a_{B/\mu} &= (2.81 \pm 0.40) \times 10^{-3},
\label{Bresults} \\
b_{B/\mu} &= (4.14 \pm 0.55) \times 10^{-6},
\label{Bresults2}\\
R^2_{B/\mu} &= 0.92.
\end{align}
\end{subequations}

\begin{figure} [t]
\includegraphics[width=\columnwidth]{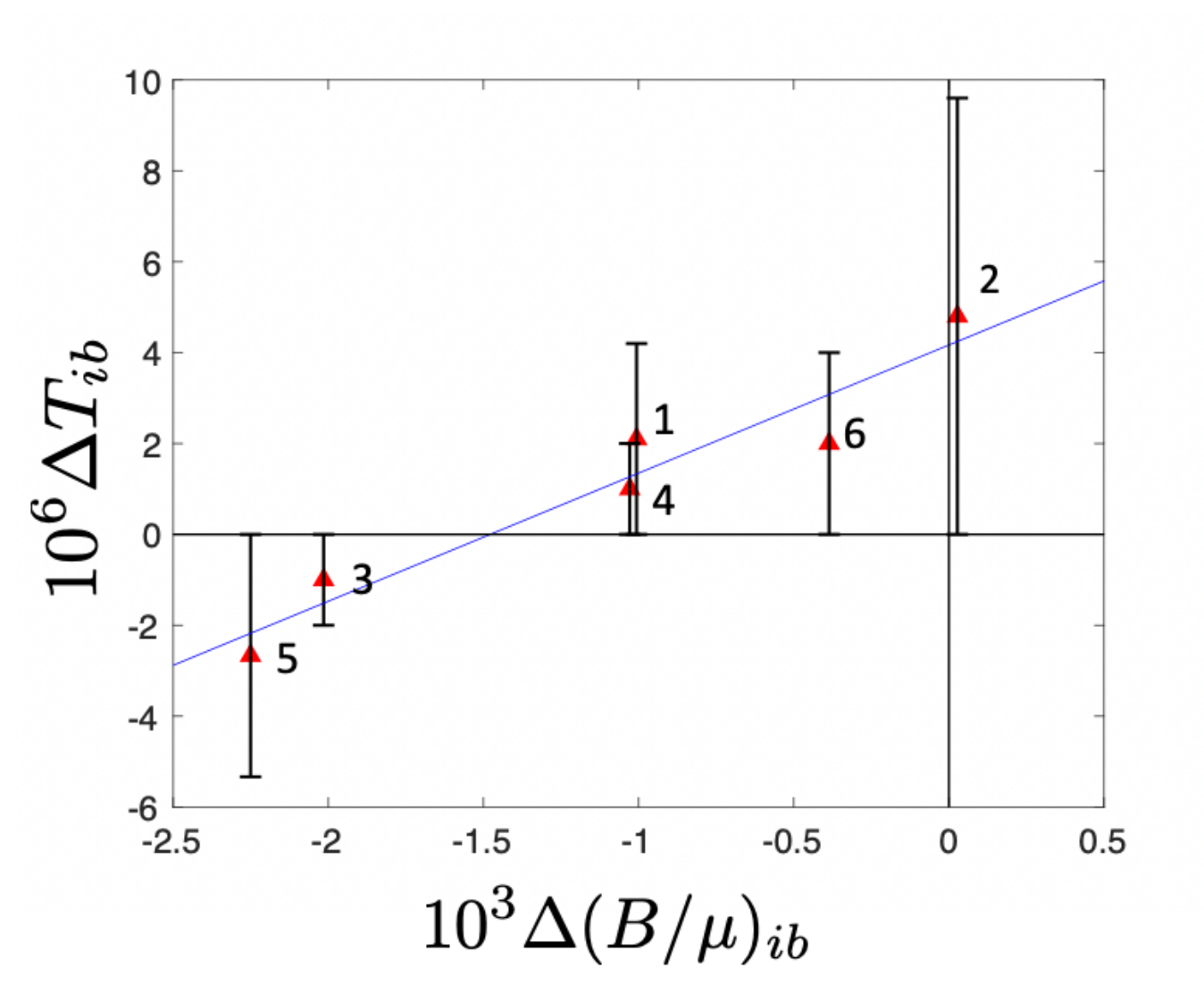}
\caption{Plot of the 6 samples tabulated in Table \ref{Potter_results} measured by Potter compared to brass with artificially assigned error bars that are consistent with a null effect.}
\label{inflatedError}
\end{figure}

Next, for illustrative purposes, we assigned an uncertainty to each data point that is consistent with a null measurement, as shown in Fig.~\ref{inflatedError}.
A weighted least squares fit with these assumed uncertainties results in a slope and intercept give by
\begin{subequations}
\begin{align}
a_{B/\mu} = (2.29 \pm 1.0) \times 10^{-3},
\label{leastsquares} \\
b_{B/\mu} = (3.46 \pm 1.5) \times 10^{-6}.
\label{leastsquares2}
\end{align}
\end{subequations}
We note that the slope and the intercept of the lines in Fig. \ref{inflatedError}  and Fig. \ref{6points} agree.  If the simple model given by Eq.~(\ref{GeneralPeriod}) is correct, the positive slope indicates that the directional angle of $\vec{\mathbb{F}}_{5}$ satisfies $|\phi|<0$, i.e., it has a component that is vertically upward.  Of course, with uncertainties for each data point, the uncertainties of the slope and intercept are increased compared to Eqs. (\ref{Bresults}) and (\ref{Bresults2}); however, it is important to emphasize that the slope still represents a non-zero effect, and the value of $R^{2}$ indicates a  reasonably good fit. 

However, the intercepts of the lines in Figs. \ref{6points} and \ref{inflatedError} are  {\em not} consistent with zero, within the uncertainties, as shown by Eqs.~(\ref{Bresults2}) and (\ref{leastsquares2}).
According to Eq. (\ref{GeneralPeriod}), a fifth-force contribution would not lead to  an intercept.  (We note that a similar analysis for the EPF experiment gives a vanishing intercept \cite{EPF paper}.)
The  non-vanishing intercept suggests the presence of a non-baryonic systematic effect, which could arise from an unanticipated experimental systematic effect or in one of the multiplicative systematic corrections Potter made, as discussed earlier.  Alternatively, it is important to note that, due to Potter's experimental procedure, all of the period differences are taken with an experiment group's brass sample.  A systematic involving these brass samples would also lead to a non-vanishing intercept.

\subsection{Alternative Couplings}
\begin{figure} [t]
\includegraphics[width=\columnwidth]{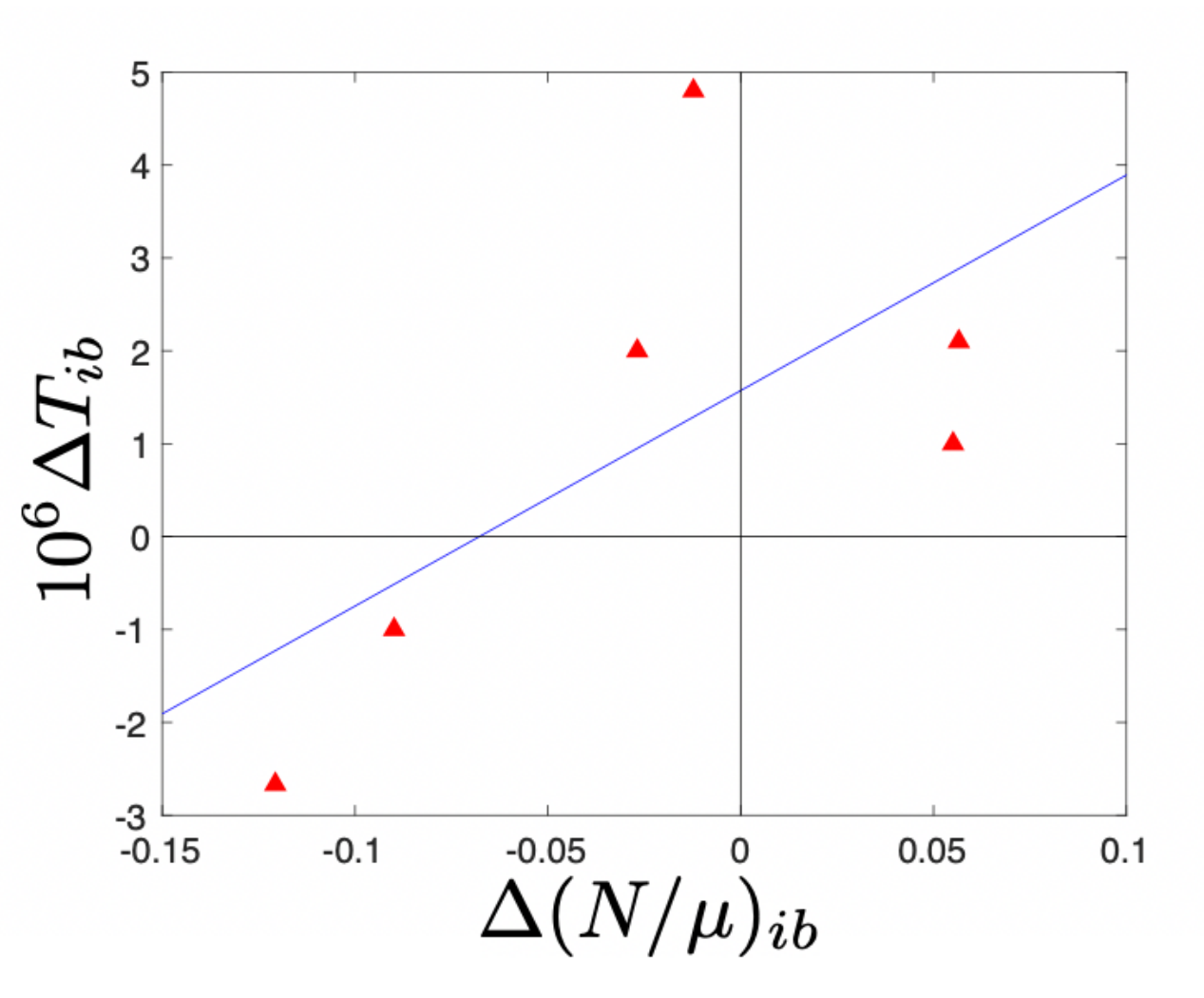}
\caption{Search for a correlation of the data with $\Delta N/\mu$ with a plot of the 6 samples  tabulated in Table \ref{summary}.}
\label{NvsT}
\end{figure}

\begin{figure} [h!]
\includegraphics[width=\columnwidth]{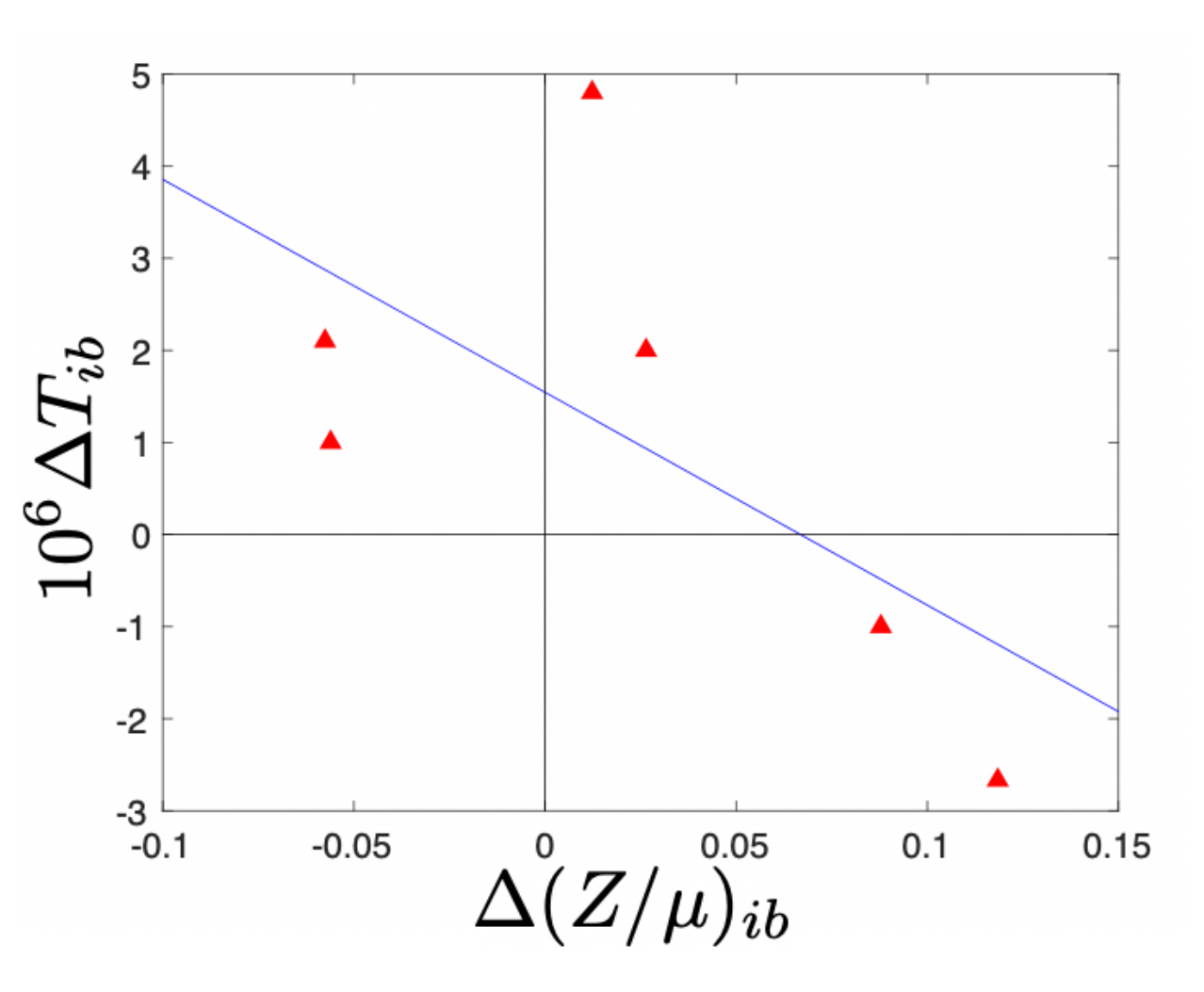}
\caption{Search for a correlation of the data with $\Delta Z/\mu$ with a plot of the 6 samples  tabulated in Table \ref{summary}.}
\label{ZvsT}
\end{figure}

The inference from Figs. \ref{6points} and \ref{inflatedError} of a correlation between baryon number $B$ and the periods measured by Potter can be further strengthened by comparing those results with  other possible correlations such as with neutron number $N$ and atomic number $Z$.   A summary of the composition values used for the samples used by Potter is given in Table \ref{summary}.  In Figs. \ref{NvsT} and \ref{ZvsT}, we plot the analogs of Fig. \ref{6points} for the variables $N/\mu$, and $Z/\mu$, where $N=B-Z$.

\begin{table}[]
\caption{Summary of Data for the Potter Samples. (The mahogany datum is not included for reasons discussed in the text.)}
\footnotesize
\begin{tabular}{crrrr}
\hline
Index ($i$) & $ 10^6 \Delta T_{ib}$ & $10^3 \Delta(B/\mu)_{ib}$ & $ 10^2 \Delta(N/\mu)_{ib}$ & $10^2 \Delta(Z/\mu)_{ib}$ \\ \hline
1        	& 2.100            		& $-$1.001       				& 5.655  					& $-$5.755  \\ 
2        	& 4.800            		& 0.047           				& $-$1.225   				& 1.230	\\
3        	& $-$1.000       		& $-$2.009     				& $-$8.987     				& 8.786 	\\ 
4        	& 1.000            		& $-$1.023       				& 5.507					& $-$5.609	\\ 5        	& $-$2.667       		& $-$2.245      				& $-$12.07       				& 11.84 	\\ 
6        	& 2.000            		& $-$0.380       				& $-$2.682    				& 2.644		\\ \hline
\end{tabular}
\label{summary}
\end{table}

 From Fig.~\ref{NvsT},  the slope, intercept, and coefficient of determination for a fifth force  coupling linearly to $N$ from the linear regression give, 
\begin{subequations}
\begin{align}
a_{N/\mu} & = (2.32 \pm 1.4) \times 10^{-5}, 
\label{Nresult} \\
b_{N/\mu} & = (1.57 \pm 0.96) \times 10^{-6},\\
R^2_{N/\mu} &= 0.42 .
\end{align}
\end{subequations}
Similarly, from Fig.~\ref{ZvsT}, the slope, intercept, and coefficient of determination for a fifth force  coupling linearly to $Z$ are given by
\begin{subequations}
\begin{align}
a_{Z/\mu} & = (-2.31 \pm 1.4) \times 10^{-5}, \\
b_{Z/\mu} & = (1.54 \pm 0.96) \times 10^{-6}, \\
R^2_{Z/\mu} & = 0.41 . 
\label{Zresults}
\end{align}
\end{subequations}
 It is clear from both the plots, and the associated fits in Eqs. (\ref{Nresult})--(\ref{Zresults}), that there is no significant correlation between either $\Delta(Z/\mu)$ nor $\Delta(N/\mu)$ and the periods in the Potter data. The slope of both plots are non-zero within the uncertainties, but $R^2 \approx 0.4$ suggests that the difference in composition for $\Delta(Z/\mu)$ and $\Delta(N/\mu)$ does not properly predict the value for $\Delta T_{ib}$. Therefore, $\Delta(Z/\mu)$ and $\Delta(N/\mu)$ are not good predictors of a composition dependent force affecting the Potter data. This reinforces the significance of the correlation of the Potter data with $\Delta(B/\mu)$ as embodied in Eq. (\ref{GeneralPeriod}).

\section{Discussion}

Since the Potter experiment has significantly less precision than the EPF experiment, and many experiments have shown no evidence of a violation of the WEP principle \cite{Wagner,Tino, Franklin, Fischbach Book, Adelberger, Will}, it is remarkable that one finds a pattern in the Potter data very similar to what was observed in the EPF data \cite{Fischbach PRL}.  As noted in the Introduction, we have shown previously that  data from a significant number of independent pairs of samples are needed to test for the presence of a composition-dependent fifth force \cite{EPF paper}. To establish this conclusion we started with the results from the original E\"otv\"os experiment \cite{EPF, EPF Book 1, EPF Book 2}, and inflated their uncertainties to create a data set where the acceleration differences of each pair were consistent with a null effect. We then demonstrated quantitatively that, even in this extreme circumstance, the modified E\"otv\"os data set still supported the presence of a composition-dependent fifth force. 

What is new in our current presentation is that we can now support the previous conclusion with a set of  actual published experimental data obtained by Potter \cite{Potter}. Although his period differences are consistent with a null effect, as seen in Fig. \ref{inflatedError} the resulting slope of the line provides a clear example of how a fifth force signal can nonetheless be embedded in the \emph{pattern} formed by a set of measured quantities.
These results reinforce the argument presented in Ref.~\cite{EPF paper} calling for a new round of high precision tests of the WEP using many multiple samples.   We also hope that the result of this reanalysis of the Potter experiment is an important clue that  will point to a final resolution of the E\"otv\"os paradox.

\section*{Acknowledgments}

We thank Virginia Mills, Beth Lindsay, and Jeff Beck for their assistance in obtaining a copy of Harold Potter's original manuscript from the archives of the Royal Society.  We also thank Eva Haviarova and Mike Jenkins from the Purdue Forestry Department for information on the chemical composition of mahogany, as well as Laura Cayon and Arianna Meenakshi McNamara for their help on the statistical analysis of the Potter data.

\pagebreak


\begin{thebibliography}{99}

\bibitem{Fischbach Book} E. Fischbach, C. Talmadge, The Search for Non-Newtonian Gravity, Springer, 1999.

\bibitem{Adelberger} E. G. Adelberger, J. H. Gundlach, B. R. Heckel, S. Hoedl, and S. Schlamminger,  Prog. Part. Nucl. Phys. 62  (2009) 102.

\bibitem{Damour and Donoghue}  T. Damour and J. F. Donoghue, Phys. Rev. D 82 (2010) 084033.

\bibitem{Jaeckel} J. Jaeckel, A. Ringwald, Annu. Rev. Nucl. Part. Sci. 60 (2010) 405.

\bibitem{Damour}  T. Damour, Class. Quantum Grav. 29 (2012) 184001.

\bibitem{Will}  C. M. Will, Theory and Experiment in Gravitational Physics, 2nd ed., Cambridge University Press, 2018.

\bibitem{Tino} G. M. Tino et al., Prog. Part. Nucl. Phys. 112 (2020) 103772.

\bibitem{Fayet 1996} P. Fayet, Class. Quantum Grav. 13 (1996) A19.

\bibitem{Adelberger 2007} E. G. Adelberger, B. R. Heckel, S. Hoedl, C. D. Hoyle,  D. J. Kapner, A. Upadhye, Phys. Rev. Lett. 98 (2007) 131104.

\bibitem{Dent}  T. Dent, Phys. Rev. Lett. 101 (2008) 041102.

\bibitem{Kraiselburd} L. Kraiselburd, H. Vucetich, Phys. Lett. B 718 (2012) 21.

\bibitem{Dobrescu}  B. A. Dobrescu, I. Mocioiu, JHEP 11 (2006) 005.

\bibitem{Fadeev} P. Fadeev, Y. V. Stadnik, F. Ficek, M. G. Kozlov, V. V. Flambaum, D. Budker, Phys. Rev. A 99 (2019) 022113.

\bibitem{EPF paper}  E. Fischbach, J. T. Gruenwald, D. E. Krause, M. H. McDuffie, M. J. Mueterthies, C. Y. Scarlett, Phys. Lett. A 399 (2021) 127300; arXiv:2012.02862 [hep-ph] (2020).

\bibitem{Newton} I. Newton, The Principia, University of California Press, 1999.

\bibitem{Bessel} F. Bessel, Ann. Phys. (Leipzig) 101 (1832) 401.

\bibitem{EPF}  R. V. E\"{o}tv\"{o}s, D. Pekár, and E. Fekete, Ann. Phys. (Leipzig) 68 (1922) 11.

\bibitem{EPF Book 1} Z. Szab\'{o}, ed., Three Fundamental Papers of Lor\'{a}nd E\"{o}tv\"{o}s, E\"{o}tv\"{o}s Lor\'{a}nd Geophysical Institute of Hungary,1998.

\bibitem{EPF Book 2}  \'{E}. Kil\'{e}nyi, The E\"{o}tv\"{o}s Experiment in its Historical Context, Unicus M\"{u}hely, 2019.

\bibitem{Potter} H. Potter, Proc. R. Soc. A 104 (1923) 588

\bibitem{Potter manuscript} H. Potter, unpublished manuscript (1923).

\bibitem{Dicke} P. G. Roll, R. V. Krotkov, R. H. Dicke, Ann. Phys. (NY) 26 (1964) 442.

\bibitem{Braginskii}  V. B. Braginskii, V. I. Panov, Sov. Phys. JETP 34 (1972) 463.

\bibitem{Wagner} T. A. Wagner, S. Schlamminger, J. H. Gundlach, E. G. Adelberger, Class. Quantum Grav. 29 (2012) 184002.

\bibitem{Asenbaum}  P. Asenbaum, C. Overstreet, M. Kim, J. Curti, and M. A. Kasevich, Phy. Rev. Lett. 125 (2020) 191101.

\bibitem{Touboul PRL} P. Touboul et al., Phys. Rev. Lett. 119 (2017) 231101.

\bibitem{Touboul CQG} P. Touboul et al., Class. Quantum Grav. 36 (2019) 225006.

\bibitem{Fischbach PRL} E. Fischbach, D. Sudarsky, A. Szafer, C. Talmadge, S. H. Aronson,  Phys. Rev. Lett. 53 (1986) 3.

\bibitem{AoP} E. Fischbach, D. Sudarsky, A. Szafer, C. Talmadge, S. H. Aronson, Ann. Phys. (NY) 182 (1988) 1.

\bibitem{Fischbach memoir}  E. Fischbach, Eur. Phys. J. H 40 (2015) 385.

\bibitem{Franklin}  A. Franklin, E. Fischbach, The Rise and Fall of the Fifth Force, 2nd ed., Springer, 2016.

\bibitem{Fischbach PoS}  E. Fischbach, D. E. Krause, PoS (FFK2019) 039.

\bibitem{Potter2} H. Potter, Proc. R. Soc. A 113 (1927) 731.

\bibitem{Mahogany} A. Bikoro Bi Athomo, et al., Industrial Crops and Products 113 (2018) 167.

\bibitem{Duralumin} D. D. Staley, A. C. Wilbraham, M. S. Matta, Essentials of Chemistry, Benjamin/Cummings (1984) 327.

\bibitem{Claypool} A. Franklin, R. Laymon, Measuring Nothing Repeatedly, Morgan \& Claypool, 2019.

\end{thebibliography}
\end{document}